\documentclass[twocolumn,superscriptaddress,showpacs,preprintnumbers,amsmath,amssymb]{revtex4}

\usepackage{graphicx,color}
\usepackage{epsfig}
\usepackage{dcolumn}
\usepackage{bm}
\usepackage{CJK}

\newcommand{\lc}{\left<}
\newcommand{\rc}{\right>}
\newcommand{\lr}{\left|}
\newcommand{\rl}{\right|}
\newcommand{\lb}{\left(}
\newcommand{\rb}{\right)}
\newcommand{\ls}{\left[}
\newcommand{\rs}{\right]}

\newcommand{\ff}[1]{\frac{1}{#1}}

\newcommand{\bo}{\boldsymbol}

\begin{document}
\begin{CJK*}{GBK}{song}

\title{Localized form of Fock terms in nuclear covariant density functional theory}

\author{Haozhao Liang}
 \affiliation{State Key Laboratory of Nuclear Physics and Technology, School of Physics, Peking University, Beijing 100871, China}

\author{Pengwei Zhao}
 \affiliation{State Key Laboratory of Nuclear Physics and Technology, School of Physics, Peking University, Beijing 100871, China}

\author{Peter Ring}
 \affiliation{Physik Department, Technische Universit\"{a}t M\"{u}nchen, D-85747 Garching, Germany}
 \affiliation{State Key Laboratory of Nuclear Physics and Technology, School of Physics, Peking University, Beijing 100871, China}

\author{Xavier Roca-Maza}
 \affiliation{INFN, Sezione di Milano, via Celoria 16, I-20133 Milano, Italy}

\author{Jie Meng}
 \affiliation{State Key Laboratory of Nuclear Physics and Technology, School of Physics, Peking University, Beijing 100871, China}
 \affiliation{School of Physics and Nuclear Energy Engineering, Beihang University,
              Beijing 100191, China}
 \affiliation{Department of Physics, University of Stellenbosch, Stellenbosch 7602, South Africa}

\date{\today}

\begin{abstract}
In most of the successful versions of covariant density functional theory in nuclei,
the Fock terms are not included explicitly, which leads to local functionals and forms the basis of their widespread applicability at present. However, it has serious consequences for the description of Gamow-Teller resonances (GTR) and spin-dipole resonances (SDR) which can only be cured by adding further phenomenological parameters. Relativistic Hartree-Fock models do not suffer from these problems. They can successfully describe the GTR and SDR as well as the isovector part of the Dirac effective mass without any additional parameters. However, they are non-local and require considerable numerical efforts.
By the zero-range reduction and the Fierz transformation, a new method is proposed to take into account the Fock terms in local functionals, which retains the simplicity of conventional models and provides proper descriptions of the spin-isospin channels and the Dirac masses.

\end{abstract}


\maketitle

Research on quantum mechanical many-body problems is essential in many areas of modern physics.
By reducing the many-body problem formulated in terms of $N$-body wave functions to the one-body level with local density distributions $\rho(\bo r)$,
the density functional theory (DFT) of Kohn and Sham \cite{Kohn1965} has accomplished great success. No other method achieves comparable accuracy at the same computational costs.
Even though the exact form of the functionals is always extremely difficult to determine, this theory states that there exists a local single-particle potential $V_{\rm KS}(\bo r)$ so that the exact ground-state density of the interacting system can be reproduced by non-interacting particles moving in this local potential.

In nuclear physics, the DFT is used since the 1970s with great success to describe both ground states and excited states of nuclei throughout the periodic chart \cite{Bender2003}. In contrast to Coulombic systems, the spin and isospin degrees of freedom play an essential role, and the corresponding functionals contain many phenomenological parameters whose numbers can be reduced considerably by the symmetries and properties of underlying theories. Therefore, the covariant density functionals \cite{Serot1986,Ring1996,Vretenar2005,Meng2006} are of particular interest in nuclear physics. They exploit basic properties of QCD at low energies, in particular, symmetries and the separation of scales \cite{Lalazissis2004}. They provide a consistent treatment of the spin degrees of freedom, they include the complicated interplay between the large Lorentz scalar and vector self-energies induced on the QCD level by the in medium changes of the scalar and vector quark condensates \cite{Cohen1992}, and they include the nuclear currents induced by the spatial parts of the vector self-energies. Furthermore the Lorentz symmetry puts stringent restrictions on the number of parameters in the corresponding functionals without reducing the quality of the agreement with experimental data.

Over the years, a large variety of nuclear phenomena have been described successfully by these methods \cite{Vretenar2005,Meng2006,Paar2007,Niksic2011}: the equation of state in symmetric and asymmetric nuclear matter, ground-state properties of finite spherical and deformed nuclei all over the nuclear chart, collective rotational and vibrational excitations, as well as fission landscapes and low-lying spectra of transitional nuclei involving quantum phase transitions in finite nuclear systems.

Most of these successes could only be achieved because the underlying functionals depend only on local densities and currents. The non-local Fock terms of the meson exchange potentials are not taken into account explicitly and thus the theoretical framework is relatively simple and computational costs are low. However, a few common problems have been found in the widely used relativistic Hartree (RH) method in the literature,
1) it is difficult to disentangle the effects of isovector-scalar and isovector-vector channels by the nuclear ground-state properties, unless a careful tuning is performed based on selected microscopic calculations~\cite{Roca-Maza2011};
2) additional adjustment is needed for charge-exchange spin-flip excitations, such as Gamow-Teller resonances (GTR) and spin-dipole resonances (SDR);
3) nucleon-nucleon tensor interactions are practically missing so that the nuclear shell structures and their evolutions are not well reproduced.

Since a few years, relativistic Hartree-Fock (RHF) calculations are possible for spherical nuclei all over the nuclear chart~\cite{Long2006}. In contrast to the simple Hartree calculations, they reproduce successfully the effective mass splitting~\cite{Long2006}, the spin-isospin resonances~\cite{Liang2008,Liang2012}, and shell structure evolutions~\cite{Long2007,Long2008}. However, since RHF theory introduces non-local potentials $V_{\rm HF}(\bo r, \bo r')$, its theoretical framework is much more involved and the computational costs are extremely heavy and the simplicity of the Kohn-Sham scheme is lost. This restricts the applicability of deformed RHF theory to a few very light nuclei~\cite{Ebran2011} and it prevents the treatment of effects beyond mean-field.

It is therefore highly desirable to stay in nuclear physics too within the conventional Kohn-Sham scheme and to find a covariant density functional based on only local potentials, yet keeping the merits of the exchange terms. In this paper, we start from the important observation that one of the successful and most widely used RHF parameterizations PKO2~\cite{Long2008} includes only three relatively heavy mesons, $\sigma$, $\omega$, and $\rho$,  but no pions. Since the masses of these mesons are heavy, the zero-range reduction becomes a reasonable approximation. Moreover, with the help of Fierz transformation~\cite{Okun1982,Sulaksono2003}, the Fock terms can be expressed as local Hartree terms.
The main goal here is to investigate the validity of the local approximation to the exchange terms of the RHF theory. The isovector effective mass splitting and spin-isospin resonances will be focused, where the effect of the exchange terms is known to be significant \cite{Long2006,Liang2008,Liang2012}.

\begin{widetext}
The basic ansatz of RHF theory is an effective Lagrangian density~\cite{Bouyssy1987,Long2006},
\begin{eqnarray}
   \mathcal{L} &=& \bar\psi \ls i\gamma^\mu\partial_\mu -M -g_\sigma\sigma
   - \gamma^\mu\lb g_\omega\omega_\mu + g_\rho
   \vec\tau\cdot\vec\rho_\mu + e {\frac{1-\tau_3}{2}}
   A_\mu\rb\rs \psi\nonumber\\
   &&+\ff2\partial^\mu\sigma\partial_\mu\sigma -\ff2m_\sigma^2\sigma^2-\ff4\Omega^{\mu\nu} \Omega_{\mu\nu} +\ff2
   m_\omega^2\omega_\mu\omega^\mu -\ff4\vec R^{\mu\nu}\cdot\vec R_{\mu\nu}+\ff2m_\rho^2
   \vec\rho^\mu\cdot\vec\rho_\mu-\ff4 F^{\mu\nu}F_{\mu\nu},
\end{eqnarray}
in which nucleons are described by Dirac spinors that interact with each other by exchanging mesons and photons.
Here, all the symbols have the same meanings as in Ref.~\cite{Meng2006}.
With the Hartree-Fock and no-sea approximations, the system energy functional can be written as~\cite{Liang2009}
\begin{eqnarray}
    E = \lc \Phi_0 \rl \hat H \lr \Phi_0 \rc
    = \sum^{\rm occ}_a \lc a \rl{\mathbf{\alpha}\cdot\mathbf{p}+\beta M}\lr a \rc
        +\ff2\sum^{\rm occ}_{ab}\lc{ab}\rl{V(1,2)}\lr{ba}\rc-\ff2\sum^{\rm occ}_{ab}\lc{ab}\rl{V(1,2)}\lr{ab}\rc,
\end{eqnarray}
\end{widetext}
where $\lr \Phi_0 \rc$ is the ground-state wave function as a Slater determinant, $a,b$ represent the occupied single-particle states, and the two-body interaction $V(1,2)$ includes the following meson-nucleon interactions,
\begin{eqnarray}
    V_\sigma(1,2) &=& -[g_\sigma\gamma_0]_1[g_\sigma\gamma_0]_2D_\sigma(1,2),\label{sigma}\nonumber\\
    V_\omega(1,2) &=& [g_\omega\gamma_0\gamma^\mu]_1[g_\omega\gamma_0\gamma_\mu]_2D_\omega(1,2),\label{omega}\\
    V_\rho(1,2) &=& [g_\rho\gamma_0\gamma^\mu\vec\tau]_1\cdot [g_\rho\gamma_0\gamma_\mu\vec\tau]_2D_\rho(1,2),\label{rho}\nonumber
\end{eqnarray}
and the photon-nucleon interactions.
The corresponding finite-range Yukawa propagators of the mesons read
\begin{equation}\label{Eq:Yukawa}
    D_i(\bo r_1, \bo r_2) = \ff{4\pi}\frac{e^{-m_i|\bo r_1-\bo r_2|}}{|\bo r_1-\bo r_2|}~~~\mbox{or}~~~D_i(\bo q) = \ff{m_i^2+\bo q^2}
\end{equation}
with $i = \sigma, \omega, \rho$.

At the scale of nuclear structure, the degrees of freedom that have to be taken into account explicitly are pions and nucleons.
The exchange of heavy mesons associated with short-distance dynamics cannot be resolved at low energies that characterize nuclear binding and can be represented by local four-point nucleon-nucleon interactions.
This type of approaches has been demonstrated to be suitable for the descriptions of the nuclear ground-state and excited state properties~\cite{Niksic2008,Zhao2010}.
In the heavy meson limit, $m_i\gg q$, the Yukawa propagator in Eq.~(\ref{Eq:Yukawa}) can be expanded as
\begin{equation}\label{Eq:DZR}
    D_i(\bo q) = \ff{m_i^2} - \frac{\bo q^2}{m_i^4} + \cdots~~~\mbox{or}~~~D_i(\bo r_1, \bo r_2) \approx \ff{m_i^2}\delta(\bo r_1-\bo r_2)
\end{equation}
within the zero-order approximation. In next order, one finds the gradient terms,
which are important for the description of ground-state properties like nuclear radii. They are not taken into account in the present calculations of nuclear matter and collective excitations, because the zero-order approximation is accurate enough for the physics discussed here.
The zero-range coupling strengths $\alpha^{\rm HF}$ related to the meson-nucleon coupling strengths and meson masses read
\begin{equation}\label{Eq:aZR}
    \alpha^{\rm HF}_S = -\frac{g_\sigma^2}{m_\sigma^2},\qquad
    \alpha^{\rm HF}_V = \frac{g_\omega^2}{m_\omega^2},\qquad
    \alpha^{\rm HF}_{tV} = \frac{g_\rho^2}{m_\rho^2}.
\end{equation}
In present paper, the subscripts $S$, $V$, $T$, $PS$, and $PV$ label the scalar, vector, tensor, pseudoscalar, and pseudovector channels, respectively, while without (with) $t$ indicates the isoscalar (isovector) nature.

According to the Fierz transformation~\cite{Okun1982,Sulaksono2003}, the exchange terms can be expressed as linear superpositions of direct terms with the equivalent coupling strengths
\begin{equation}\label{Eq:alpha}
    \alpha^{\rm H}_j = \sum_k c_{jk} \alpha^{\rm HF}_k,
\end{equation}
where $j=\{S,tS,V,tV,T,tT,PS,tPS,PV,tPV\}$, $k=\{S, V, tV\}$, and the transpose of matrix $C$ reads
\begin{equation*}
    C^T = \ff{16}
    \ls \begin{array}{rrrrrrrrrr}
        14  & -2 & -2 & -2 & -1 & -1 & -2 & -2 & 2  & 2 \\
        -8  & -8 & 20 & 4  & 0  & 0  & 8  & 8  & 4  & 4 \\
        -24 & 8  & 12 & 12 & 0  & 0  & 24 & -8 & 12 & -4
    \end{array}\rs.
\end{equation*}

\begin{figure}
\includegraphics[width=8cm]{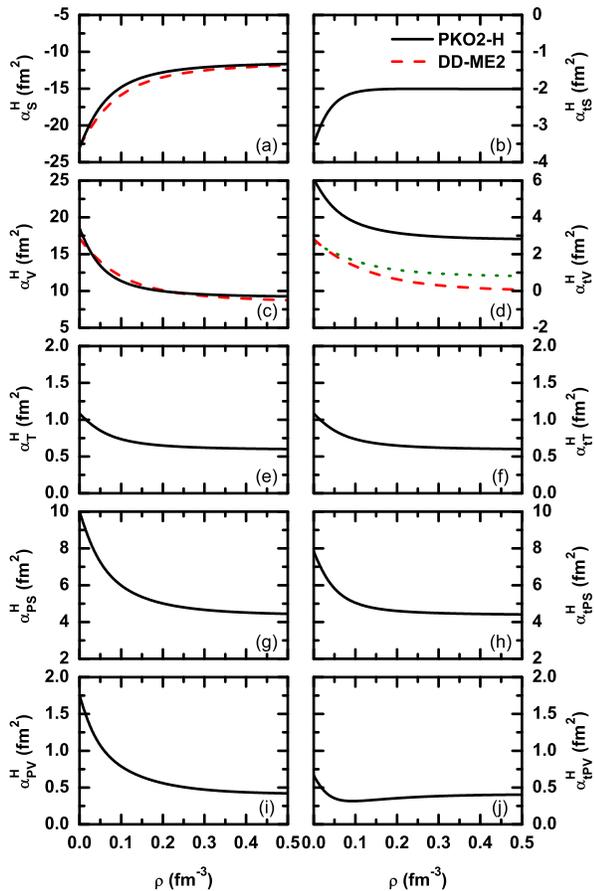}
\caption{(Color online) Zero-range coupling strengths of PKO2-H in the isoscalar (left column) and isovector (right column) channels as a function of the baryonic density. DD-ME2~\cite{Lalazissis2005} parameters are shown for comparison. The dotted line is $\alpha^{\rm H}_{tS}+\alpha^{\rm H}_{tV}$.
\label{Fig1}}
\end{figure}

In this way, the coupling strengths in ten channels are determined by the three density-dependent RHF parameters of PKO2.
The RHF equivalent parametrization thus obtained will be called PKO2-H, and its zero-range coupling strengths are shown as a function of the baryonic density in Fig.~\ref{Fig1}.
They are compared with the corresponding strength parameters of the widely used density-dependent RH parametrization DD-ME2~\cite{Lalazissis2005}, which includes only $S$, $V$, and $tV$ channels.
In the following, we denote full Hartree-Fock results with parametrization PKO2 by RHF, the present results with PKO2-H by RHF-loc, and the conventional Hartree results with parametrization DD-ME2 by RH.
Very good agreement between both sets are found for the size and the density dependence of the two isoscalar channels $S$ and $V$.
This is not surprising because in the covariant framework the balance between these two channels is very delicate and plays an essential role in both mean-field and spin-orbit potentials.

However, a substantial difference between RHF-loc and RH appears in the isovector channels. In DD-ME2 the isovector-scalar $\delta$-meson is not explicitly included and the isovector-vector $\rho$-meson determines the full isospin dependence. Eq.~(\ref{Eq:alpha}) shows that the $tV$ term in PKO2-H is strongly repulsive. It is modified by the strongly attractive $tS$ term and for the large components we find as in the isoscalar case a delicate balance between scalar and vector channels. This leads to
$\alpha^{\rm DD-ME2}_{tV}\approx\alpha^{\rm H}_{tS}+\alpha^{\rm H}_{tV}$ (see the dotted line in Fig.~\ref{Fig1}).

\begin{figure}
\includegraphics[width=8cm]{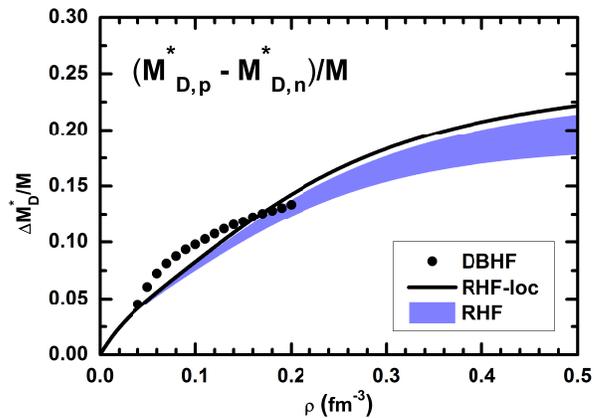}
\caption{(Color online) Dirac mass splitting between neutron and proton in neutron matter for RHF (shadow band) and RHF-loc (solid line) as the function of the baryonic density.
The predictions of the Dirac-Brueckner-Hartree-Fock calculation~\cite{Dalen2007} are shown as close symbols.
\label{Fig2}}
\end{figure}

The scalar fields in the isovector channel determine the Dirac mass splitting between neutron and proton in asymmetric nuclear matter. Due to the absence of the $tS$ channel, it vanishes identically for RH. This contradicts to the results of microscopic Dirac-Brueckner-Hartree-Fock (DBHF) calculations~\cite{Dalen2007}. In Fig.~\ref{Fig2}, the Dirac mass splittings in neutron matter are presented for RHF-loc (solid line) and DBHF~\cite{Dalen2007} (close symbols). The corresponding cuves are compared also with results for the full RHF calculations. In RHF, the Dirac mass is momentum dependent and therefore as a shadow band is shown, where its upper (lower) border corresponds to the results calculated with neutrons at zero (Fermi) momentum. Note that the proton Fermi momentum vanishes in neutron matter. In the present RHF equivalent calculations, the Dirac mass does not depend on the momentum, but it maintains the characteristic behavior of the RHF results. It is also in very good agreement with the microscopic DBHF results. This indicates the constraints introduced by the Fock terms of the RHF scheme into the $tS$ channel of the present density functional are straight forward and quite robust.

In panels (e)--(j) of Fig.~\ref{Fig1}, it is found that the remaining six channels of PKO2-H do not vanish. They are explicitly determined in Eq.~(\ref{Eq:alpha}) by the exchange effects of the RHF scheme.
In contrast, the coupling strengths $\alpha^{\rm H}_{(t)PS}$ and $\alpha^{\rm H}_{(t)PV}$ are practically out of control in usual fitting processes since they are absent in the description of ground states due to parity conservation. However they play a crucial role in the properties of charge-exchange modes.

In the following, we demonstrate their importance for Gamow-Teller and spin-dipole resonances excited by charge-exchange reactions in spherical nuclei. Based on the unperturbed single-particle spectra calculated by RHF theory, we perform calculations based on the proton-neutron random phase approximation (RPA) using the particle-hole (\textit{ph}) interactions
\begin{align}
\label{Eq:ph}
      V_{tS}(1,2) &= \alpha^{\rm H}_{tS}[\gamma_0\vec\tau]_1\cdot[\gamma_0\vec\tau]_2\delta(\bo r_1-\bo r_2),\nonumber\\
      V_{tV}(1,2) &= \alpha^{\rm H}_{tV}[\gamma_0\gamma^\mu\vec\tau]_1\cdot[\gamma_0\gamma_\mu\vec\tau]_2\delta(\bo r_1-\bo r_2),\nonumber\\
      V_{tT}(1,2) &= \alpha^{\rm H}_{tT}[\gamma_0\sigma^{\mu\nu}\vec\tau]_1\cdot[\gamma_0\sigma_{\mu\nu}\vec\tau]_2\delta(\bo r_1-\bo r_2),\\
      V_{tPS}(1,2) &= \alpha^{\rm H}_{tPS}[\gamma_0\gamma_5\vec\tau]_1\cdot[\gamma_0\gamma_5\vec\tau]_2\delta(\bo r_1-\bo r_2),\nonumber\\
      V_{tPV}(1,2) &= \alpha^{\rm H}_{tPV}[\gamma_0\gamma_5\gamma^\mu\vec\tau]_1\cdot[\gamma_0\gamma_5\gamma_\mu\vec\tau]_2\delta(\bo r_1-\bo r_2),\nonumber
\end{align}
while all isoscalar contributions vanish.

\begin{figure}
\includegraphics[width=8cm]{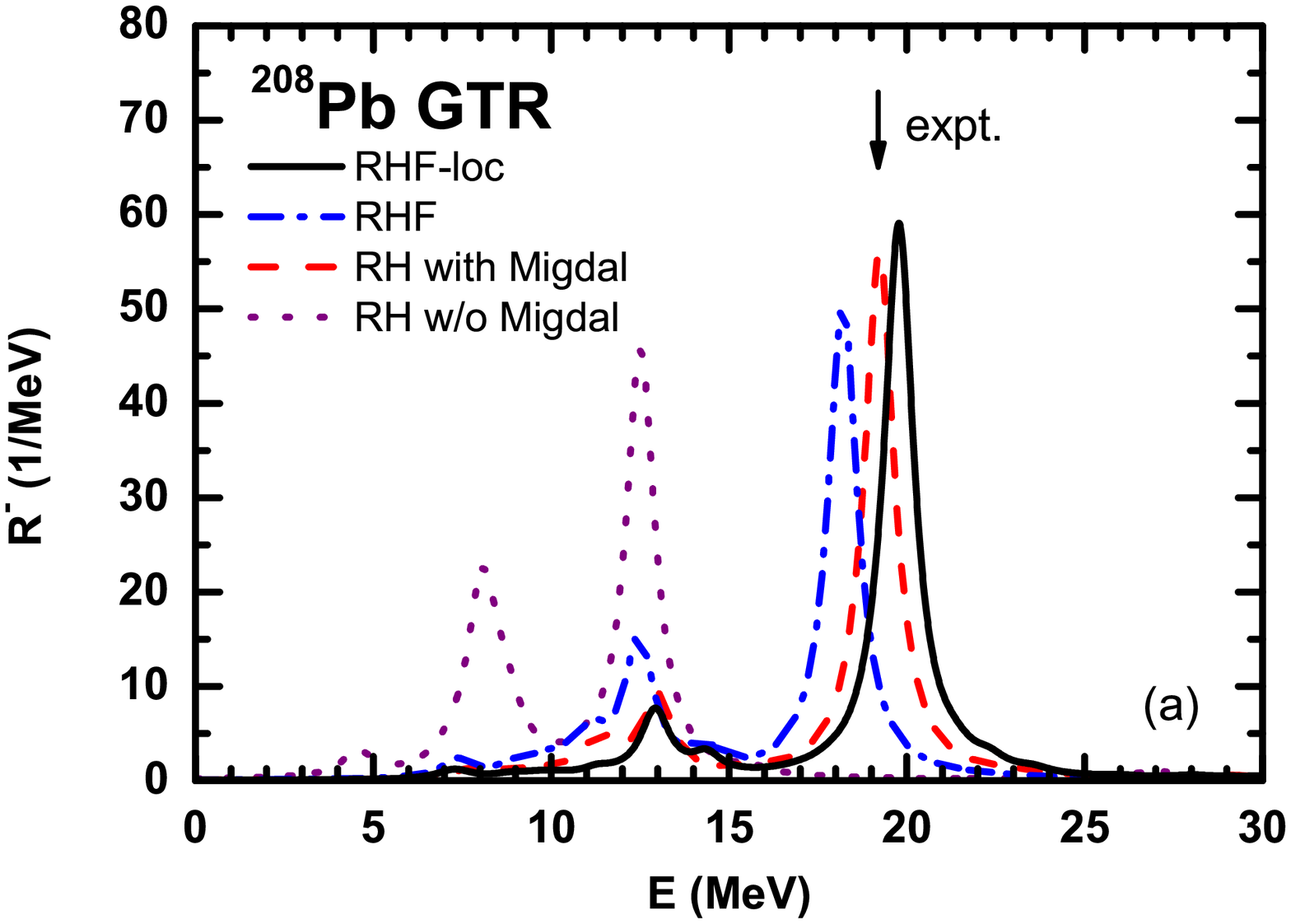}\\
\includegraphics[width=8cm]{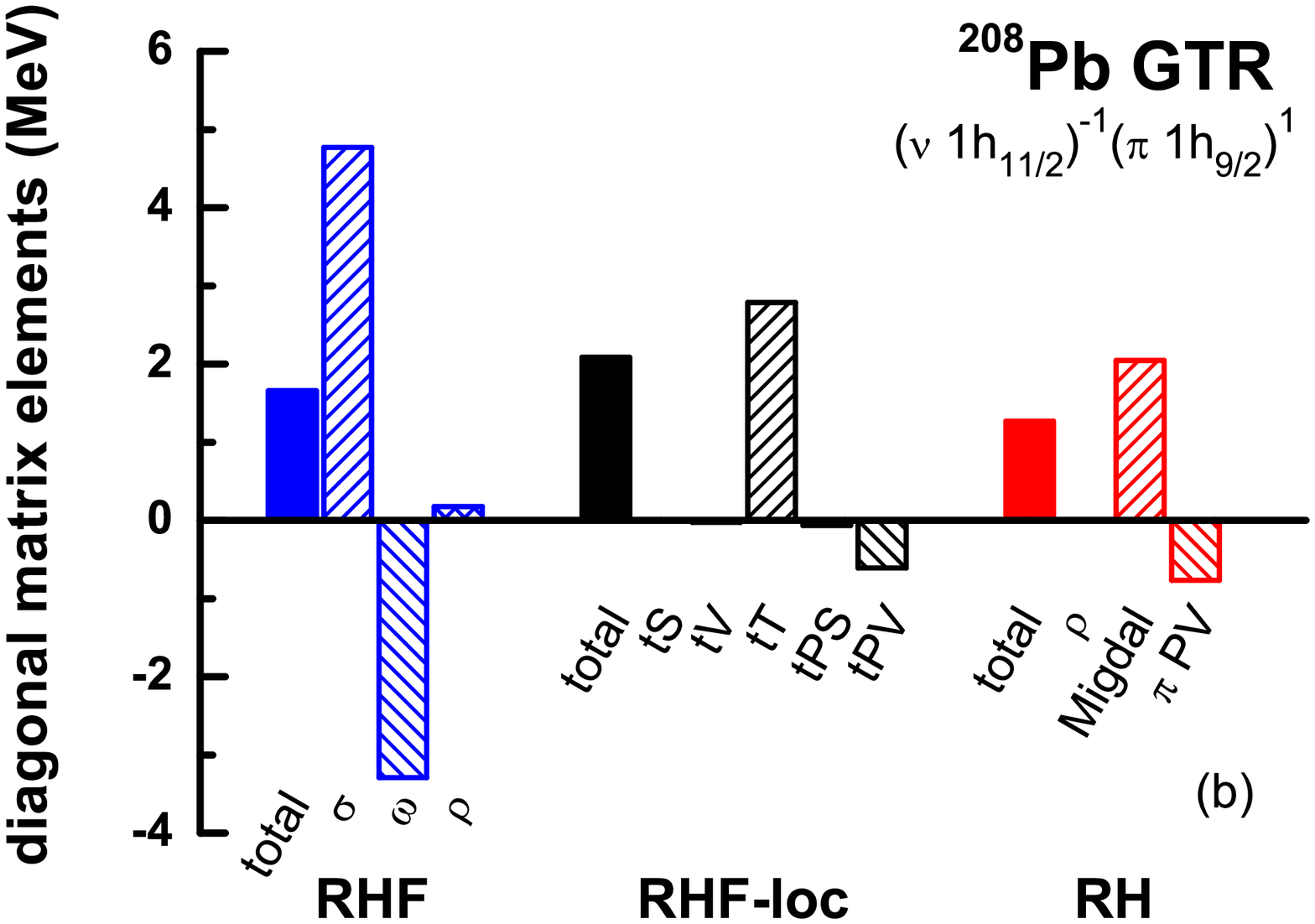}
\caption{(Color online) Upper panel: Strength distribution of the GTR in $^{208}$Pb calculated with a Lorentzian smearing parameter of $1$~MeV.
The experimental peak energy~\cite{Akimune1995} is denoted by an arrow.
Lower panel: Diagonal matrix elements of particle-hole residual interactions for the $(\nu 1h_{11/2})^{-1}(\pi 1h_{9/2})^1$ configuration. Further details are given in the text.
    \label{Fig3}}
\end{figure}

In the upper panel of Fig.~\ref{Fig3}, the GTR strength distribution in $^{208}$Pb for the RHF equivalent RPA is compared with the results of RHF+RPA and RH+RPA~\cite{Paar2004}.
For RH+RPA, a phenomenological Migdal term had to be introduced in order to reproduce the GTR data in $^{208}$Pb \cite{Paar2004}. Here the results with and without this Migdal term are shown.
The present RPA calculations lead to quite similar results as those by the original RHF+RPA~\cite{Liang2008} calculations.
The slight difference is due to the zero-range approximation in Eq.~(\ref{Eq:DZR}). The black arrow indicates the experimental peak energy~\cite{Akimune1995}. It is reproduced very well for RHF and RHF-loc, and it is fitted for RH by adjusting the Migdal term.

To understand the physical mechanisms, we present in the lower panel of Fig.~\ref{Fig3} the diagonal matrix elements of an important GT \textit{ph} configuration $(\nu 1h_{11/2})^{-1}(\pi 1h_{9/2})^1$. The total matrix element (full) is decomposed in various ways: For the original RHF+RPA~\cite{Liang2008} calculations, the total \textit{ph} strengths is essentially determined by the delicate balance between the $\sigma$- and $\omega$-mesons contributing via the exchange term to the isovector channel.  The original $\rho$-meson gives only a minor contribution. The effects of $\sigma$- and $\omega$-mesons are repulsive and attractive, respectively, since they contribute only via the Fock terms.

A very different decomposition is shown for RHF-loc, where the matrix elements in the five channels ($tS,tV,tT,tPS,tPV$) are compared. Of course, according to Eq.~(\ref{Eq:alpha}), all three mesons contribute here. It is seen that the major contribution comes from the repulsive $tT$ channel. It is reduced by the $tPV$ channel.  This can be understood by the following analysis. For the charge-exchange spin-flip modes, the dominant residual interactions are represented by the matrix elements of the operator $[\bo\sigma\vec\tau]\cdot[\bo\sigma\vec\tau]$ in the large components of Dirac spinors. From Eqs.~(\ref{Eq:ph}), one can readily find that they are composed of $[\sigma^{ij}\vec\tau]\cdot[\sigma_{ij}\vec\tau]\propto 2\alpha^{\rm H}_{tT}$ and $[\gamma_5\gamma^{i}\vec\tau]\cdot[\gamma_5\gamma_{i}\vec\tau]\propto-\alpha^{\rm H}_{tPV}$.
This decides the signs of these contributions and their amplitudes roughly differing by a factor of $4$. Furthermore, the net contribution is then approximately proportional to
\begin{equation}
    2\alpha^{\rm H}_{tT}-\alpha^{\rm H}_{tPV}
    = - \ff3\lb \alpha^{\rm H}_{S}+\alpha^{\rm H}_{V}\rb,
\end{equation}
according to the intrinsic relations in Eq.~(\ref{Eq:alpha}).
In other words, the total \textit{ph} strengths are determined by nothing but the delicate balance between the $S$ and $V$ channels (corresponding to the $\sigma$- and $\omega$-mesons), which is well established in nuclear covariant density functionals.
Therefore, the constraints introduced by the Fock terms of the RHF scheme into the \textit{ph} residual interactions are also straight forward and quite robust.

In the case of RH, the contributions from the $\rho$-meson, the Migdal term, and the pseudovector coupling of the pion are shown. It is found by fitting to the data that the free pion residual interaction is attractive and the Migdal term is repulsive.

\begin{figure}
\includegraphics[width=8cm]{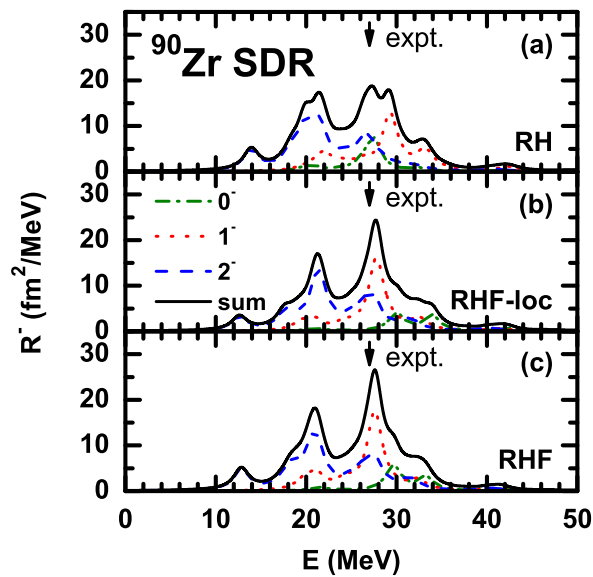}
\caption{(Color online) Strength distributions of the SDR in $^{90}$Zr calculated by (a) RH, (b) RHF-loc, and (c) RHF with a Lorentzian smearing parameter of $2$~MeV. The dash-dotted, dotted, dashed lines show the $J^\pi=0^-, 1^-, 2^-$ contributions, respectively, while the solid lines show their sum. The experimental peak energy~\cite{Yako2006} is denoted by the arrows.
\label{Fig4}}
\end{figure}

The local representation of the Fock terms can be further examined by another important charge-exchange spin-flip mode, the spin-dipole resonance (SDR). In Fig.~\ref{Fig4}, the strength distributions of SDR in $^{90}$Zr of the RHF equivalent RPA are compared with the results of RHF+RPA and RH+RPA with the Migdal term. The experimental peak energy~\cite{Yako2006} can be excellently reproduced by the original RHF+RPA and the present RHF equivalent RPA calculations. However, the traditional RH+RPA calculations present a more fragmented structure due to the poorly adjusted isovector \textit{ph} residual interactions~\cite{Marketin2012}. This is remarkably improved in the present RHF equivalent functional. Therefore, not only the total strengths but also the individual contribution from different spin-parity $J^\pi$ components are almost identical to the corresponding RHF+RPA results. In particular, this also applies for the energy hierarchy $E(2^-)<E(1^-)<E(0^-)$.

The above investigations on charge-exchange excitations illuminate that the success of the RHF+RPA approach has a solid foundation. The essential roles played by the $\sigma$- and $\omega$-mesons via the Fock terms can be successfully folded into the equivalent interactions in the Hartree level. This achievement makes the theoretical framework much simpler and the computations much more economical.

In summary, a new local RHF equivalent covariant density functional is proposed.
By the zero-range reduction and the Fierz transformation, the constraints introduced by the Fock terms of the RHF scheme are fully taken into account. In this way, the advantages of existing RH functionals can be maintained, while the problems in the isovector channel can be solved. This opens a new door for the development of nuclear local covariant density functionals with proper isoscalar and isovector properties in the future. As a consequence, the results of this investigation point in the direction of the conventional Kohn-Sham picture to represent the non-local exchange terms in the form of local potentials.

%
The authors are grateful to Ying Chen for helpful discussions. This work was partially supported by the Major State 973
Program 2007CB815000, National Natural Science Foundation of China under Grant Nos. 10975008, 11105006, 11175002, China Postdoctoral Science Foundation under Grant Nos. 20100480149, 201104031, the Research Fund for the Doctoral Program of Higher Education under Grant No. 20110001110087, and the Oversea Distinguished Professor Project from Ministry of Education No. MS2010BJDX001. We also acknowledge partial support from the DFG Cluster of Excellence ``Origin and Structure of the Universe" (www.universe-cluster.de).


\end{CJK*}
\end{document}